\documentclass[prl,twocolumn, aps,floats,showpacs,nofootinbib]{revtex4}
\usepackage{graphicx}
\usepackage{epsfig}
\usepackage{pstricks}
\usepackage{pst-coil}
\usepackage{amsmath}
\usepackage{verbatim}

\def\be{\begin{equation}}
\def\ee{\end{equation}}
\def\bea{\begin{eqnarray}}
\def\eea{\end{eqnarray}}
{\newcommand{\lsim}{\mbox{\raisebox{-.6ex}{~$\stackrel{<}{\sim}$~}}}
{
\def\mpl{M_{\rm {Pl}}}
\def\gev{{\rm \,Ge\kern-0.125em V}}
\def\tev{{\rm \,Te\kern-0.125em V}}
\def\mev{{\rm \,Me\kern-0.125em V}}

\def\Tdot{\dot T}
\def\Tmax{T_{\rm {max}}}
\def\TMAX{T_{\rm {MAX}}}

\def\tosc{t_{\rm {osc}}}
\def\Treh{T_{\rm {reh}}}
\def\Tf{T_{\rm {f}}}
\def\T1{T_1}
\def\tmax{t_{\rm {max}}}
\def\tMAX{t_{\rm {MAX}}}
\def\tMAX{t_{\rm {MAX}}}
\def\treh{t_{\rm {reh}}}
\def\Hreh{H_{\rm {reh}}}
\def\HI{H_{\rm {I}}}

\def\Rmax{R_{\rm {max}}}
\def\Rosc{R_{\rm {osc}}}
\def\nG{n_{\tilde G}}
\def\YG{Y_{\tilde G}}
\def\mG{m_{\tilde G}}
\def\rhoG{\rho_{\tilde G}}
\def\greh{g_{*\rm{reh}}}
\def\Sigmatot{\Sigma_{{\rm {tot}}}}

\def\half{\frac12}

\begin{document}
\title{\bf Perturbative Reheating and Gravitino Production in Inflationary Models}
\author{Raghavan Rangarajan}
\email{raghavan@prl.res.in}
\affiliation{Theoretical Physics Division, Physical Research Laboratory,
Navrangpura, Ahmedabad 380 009, India}
\author{Narendra Sahu}
\email{n.sahu@lancaster.ac.uk}
\affiliation{Department of Physics, Lancaster University, Lancaster, LA1 4YB, UK}
\begin{abstract}
\noindent

The low reheat temperature at the end of inflation from the gravitino
bound constrains the creation of heavy Majorana neutrinos associated with
models of leptogenesis. However, a detailed view of the reheating of the
Universe at the end of inflation implies that the maximum temperature
during reheating, $\Tmax$, can be orders of magnitude higher than the final
reheat temperature.  This then allows for the production of the heavy
Majorana neutrinos needed for leptogenesis. We carry out
the complementary calculation of the gravitino production during reheating
and its dependence on $\Tmax$.
We find that the gravitino abundance generated during reheating for a quartic
potential is
comparable to the standard estimate of the abundance generated after
reheating and study its consequences for leptogenesis.

\end{abstract}
\maketitle
\section{Introduction}
\noindent

It is presumed that the early universe went through a period of
inflation and then reheated to create the radiation dominated epoch.
If nature is supersymmetric then during the process of reheating
many supersymmetric particles would have been produced, which can
have important cosmological consequences. In particular, the
production of gravitinos in the early universe and their subsequent
evolution including
decays has attracted attention. Stable gravitinos can overclose the
universe while unstable gravitinos can affect the expansion rate of
the universe during eras prior to their decay. The decay products of
unstable gravitinos can also overclose the universe or affect light
element abundances generated during nucleosynthesis. These
cosmological consequences are a function of the gravitino energy
density, $\rhoG=\mG\nG$, where $\mG$ and $\nG$ are the mass and
number density of gravitinos. In an inflationary universe, $\nG$ is
a function of the reheat temperature. Therefore, for a fixed
$\mG$, often taken to be $O(100\gev-1\tev)$, cosmological
constraints on the energy density of gravitinos provide an upper
bound on the reheat
temperature~\cite{nos,krauss,khlopovlinde,EKN,falomkinetal,
jss,ENS,ks,kletal,moroi.95,cefo}.

The number density of gravitinos is usually obtained by considering
gravitino production in the radiation dominated era following
reheating, as in
Refs.~\cite{nos,krauss,khlopovlinde,EKN,falomkinetal,jss,ENS,ks,
kletal,moroi.95,cefo}, and it is presumed that $n_{\tilde G}=0$ at
the beginning of the radiation dominated era. Gravitinos are
produced through thermal scattering and the final gravitino
abundance
is found to be proportional to the reheat temperature, $\Treh$.
$\Treh$ is the temperature of the thermal plasma at the beginning of
the radiation dominated era at $\treh$ when the inflaton field has decayed
and the energy density of the universe is dominated by
the inflaton decay products. The cosmological constraints on $\nG$
then provide an upper bound on $\Treh \lsim 10^{6 - 9}$\gev.

Such an analysis is consistent with the instantaneous decay
approximation in which one assumes that reheating is instantaneous
and therefore $\Treh$ is the maximum temperature during reheating.
However, a more detailed understanding of (perturbative) reheating
indicates that during reheating the temperature initially rises to a
maximum temperature $\Tmax$ and then falls to $\Treh$
~\cite{kolb_book,chungetal}. In fact the maximum temperature during
the course of reheating can be as high as $10^3
\,\Treh$~\cite{giudiceetal}.

Earlier works have considered whether sufficient number densities of
heavy GUT gauge and Higgs bosons, or right-handed Majorana
neutrinos, required for GUT baryogenesis or leptogenesis
respectively, can be generated with a high $\Tmax$.  They find that
leptogenesis with Majorana neutrinos of mass $\sim 10^{10}\gev$ is
feasible.  However it is then legitimate to ask if harmful
gravitinos are also produced during the course of reheating. A
priori one might expect large production with a $\Tmax$ dependent
abundance.  This could have serious implications for leptogenesis
scenarios that invoke large $\Tmax$~\cite{chungetal,delepinesarkar,
giudiceetal}.

In Ref.~\cite{rangarajansahu} we considered this issue for an inflaton
with a potential of the form $V=\half\, m^2\phi^2$ during reheating.
We find that the abundance of gravitinos produced during reheating
is 1/3 of that produced in the radiation dominated epoch.  Our
estimate for the final gravitino abundance is of the same order as 
that obtained 
in Ref.~\cite{kkm} 
using a numerical analysis 
(also for a
quadratic potential).  
Gravitino production including the reheating era
contribution has also been obtained numerically in 
Refs.~\cite{giudiceetal1,pradlersteffen2}.
In this article we investigate the gravitino
abundance generated during the course of reheating in inflationary
models with a potential $V=(\lambda/4) \phi^4$ during reheating. The
difference between the two scenarios lies in the different equations
of state for an oscillating scalar field with a $\phi^2$ and a
$\phi^4$ potential.  In the two scenarios  the oscillating scalar
field behaves like non-relativistic and relativistic matter
respectively, i.e., $\rho_\phi \propto R^{-3},\,R^{-4}$ respectively
\cite{kolb_book1}.  This affects 
the Hubble expansion rate in
the Boltzmann equation for radiation and gravitinos during reheating, 
and the source term for radiation during reheating,
and thus the
abundance of gravitinos produced by the scattering of the
thermalised radiation.

For the $\phi^4$ potential we find that the gravitino abundance
generated during reheating is about 49\% of the gravitinos produced in
the subsequent radiation dominated epoch. The gravitino abundance
generated during reheating is a function of $\Tmax$ but, as in the
quadratic potential case, the abundance can be re-expressed as a
function of $\Treh$ only.
Including the contribution from the reheating
era
and then applying the cosmological constraints on the total
gravitino abundance
lowers the upper bound on $\Treh$ by a factor of 3/2. This
does not have a serious impact on leptogenesis scenarios discussed above.

Our results in Ref.~\cite{rangarajansahu} and in this article are
valid for chaotic inflation models and for models of inflation where
one can approximate the inflaton potential during reheating by a
$\phi^2$ or $\phi^4$ term.  
However they are not valid for a
reheating scenario that includes preheating \cite{dolgkiri,tb1,kls,stb}. 
\footnote{
Refs. \cite{greeneetal,greene.99} indicate regions of parameter space for
a quartic potential for which
bosonic and fermionic preheating can be suppressed.  Furthermore, certain
parameter values for which preheating is strong produce large non-gaussianities
in the CMBR and are hence ruled out by WMAP \cite{jokinenmazumdar}.}
Gravitino production
during preheating has been considered in
Ref.~\cite{giudiceetal1,maroto.00,kallosh.00,tsujikawa.00,
nilles.01,Nilles:2001fg,
nilles&olive.01,Greene:2002ku,podolsky}.

\section{Perturbative Reheating in Inflationary Models}
We consider an inflationary model with the inflaton potential of the
form $V=(\lambda/4)\phi^4$ during reheating.
\footnote{A quartic term can generate a quadratic term as well.  We
presume that this term, or any other mass term, is small and does not 
dominate till $\phi$ has
almost decayed away.}
The inflaton field
$\phi$ starts oscillating when the inflationary epoch ends at a
cosmic time $t=t_{\rm osc}$. While oscillating the field $\phi$ decays and the
decay products thermalise,
\footnote{
By decay we refer to dissipation of the energy density of the inflaton
field due to its coupling with other species.}
 and thus reheating occurs.
\footnote{We assume that the inflaton
products thermalise quickly as discussed in Appendix A of Ref.~\cite{
chungetal}.  
Refs.~\cite{mazumdar1,mazumdar2} discuss an alternate
description of reheating in the context of the MSSM.
They argue that in the presence of large vevs for
flat directions of MSSM fields thermalisation slows down.
However if the vevs are small $(<10^{-6}\mpl)$ then there is
no effect on reheating.  Furthermore the condensates of MSSM
fields can fragment into Q-balls and the vevs can vanish in large
parts of the universe \cite{mazumdar3}.  These are the cases we
would be considering. 
}
Assuming that
the universe is reheated through the perturbative decay of the inflaton
field, the reheating picture, in general, can be described
by~\cite{kolb_book}
\bea
\dot{\rho}_r + 4 H \rho_r &=& \Gamma_\phi\rho_\phi \label{rhoreqn}\\
H^2 &=& \frac{8\pi G}{3}(\rho_r +\rho_\phi) \, ,\label{Heqn} \eea
where $\rho_r$ and $\rho_\phi$ are the energy densities of radiation and the
inflaton respectively and $\Gamma_\phi$ is the rate of dissipation of the
inflaton field energy density.
Since the equation of state for the oscillating inflaton field and
for radiation in Eq. (\ref{Heqn}) is the same we can write the rhs
of Eq. (\ref{Heqn}) as $H_I^2 (\Rosc/R)^4$, where $H_I$ is the
Hubble parameter at $\tosc$.
\be \HI= \sqrt{ \frac{8\pi}{3}
}\frac{M_I^2}{\mpl}\,, \label{HI-value} \ee
where $M_I=V_I^{1/4}$,
$V_I$ being the inflaton energy density at  $\tosc$.
Solving Eq.(\ref{Heqn}) then gives
\be R=R_{\rm osc}\left[ 2
\HI(t-\tosc)+1\right]^{1/2}\,. \label{R-reheating} \ee
(For $t\gg \tosc$, $R\sim t^{\frac 1 2}$.)

Taking $\rho_\phi= M_I^4 (\Rosc/R)^4 \exp[-\Gamma_\phi(t-\tosc)]$
in Eq. (\ref{rhoreqn}) we then get  
\be \rho_r  =
\frac{3}{8 \pi }
\mpl^2 \HI^2 
\frac{1-\rm  e^{-\Gamma_\phi (t-\tosc)}}{\left[ 2
\HI(t-\tosc)+1\right]^{2}}
\,\, .
\label{rho_r_full_t}
\ee
However to simplify our subsequent analysis,
we ignore the change in $\rho_\phi$ due to decay in Eq. (\ref{rhoreqn}), 
which is valid till $t\lsim \treh\approx\Gamma_\phi^{-1}$.
\footnote{We will assume for now that our analysis below is valid till
$\treh$ and will later discuss this assumption.}
Then 
$\rho_\phi\approx M_I^4 (\Rosc/R)^4$ and
the solution of Eq. (\ref{rhoreqn}) is given by
\bea \rho_r  &=&
\frac{3}{8 \pi }
\mpl^2 \Gamma_\phi \HI^2
\frac{t-\tosc}{\left[ 2
\HI(t-\tosc)+1\right]^{2}}
\label{rho_r_t}\\
&=& \sqrt{ \frac{3}{32 \pi}}M_I^2 \Gamma_\phi \mpl \left( \frac{R}{R_{\rm osc}} \right)^{-2}
\left[ 1- \left(\frac{R}{R_{\rm osc}} \right)^{-2}\right]
\label{rho_r}
\eea
From Eq. (\ref{rho_r}) we see that during reheating the energy density initially
increases to a
maximum value
\be
\rho_r^{\rm max}= \frac{1}{4}\sqrt{\frac{3}{32\pi}} \Gamma_\phi\mpl M_I^2\,
\ee
at $\Rmax=\sqrt2 \Rosc$.
The maximum temperature during reheating is then
\be
T_{\rm max}=0.6 g_*^{-1/4} \left( \Gamma_\phi\mpl \right)^{1/4} M_I^{1/2}\,.
\label{T-max}
\ee
Subsequently the temperature falls as $1/R^\half$ (for $t\gg\tmax$)
until the inflaton decays at $\treh$.
Once the final decay products of $\phi$ thermalise with each other through 
sufficient
interactions the radiation density becomes
\be
\rho_r^{\rm reh}=\frac{\pi^2}{30}g_* T_{\rm reh}^4\,.
\label{rho_reh}
\ee
From equations (\ref{rho_r_t}) and (\ref{rho_reh}), and assuming
$\treh\gg\tosc$, we get the reheating temperature
\be
T_{\rm reh} \approx 0.55 g_*^{-1/4} (\mpl \Gamma_\phi)^{1/2}\,.
\label{reheat-temp}
\ee
In the following we examine the production of gravitinos during reheating,
i.e.,  from $\tosc$ to $\treh$, 
and during the subsequent radiation dominated era after $\treh$, 
and discuss its consequences.

\section{Gravitino Production}
Gravitinos are produced by the scattering of the inflaton decay products;
a list
of processes is provided in, for example, Tables 1 in Refs.~\cite{EKN,moroi.95}.
The Boltzmann equation for gravitinos is given by
\be
\frac{dn_{\tilde{G}}}{dt}+3Hn_{\tilde{G}}=
\langle \Sigmatot|v|\rangle n^2\,,
\label{boltzmann}
\ee
where $n=(\zeta(3)/\pi^2)T^3$ is the number density of scatterers
($\zeta(3)=1.20206..$ is the Riemann zeta
function of 3), $\Sigmatot$ is the total scattering cross section for
gravitino production, $v$ is the relative velocity of the incoming
particles, and $\langle...\rangle$ refers to thermal averaging.
Since the gravitino lifetime is
$~10^{7-8} (100\gev/\mG)$s~\cite{EKN} decays are not relevant during the
epoch of gravitino production for gravitinos of mass $10^{2-3}\gev$.
Hence we have
not included the gravitino decay term in Eq.~(\ref{boltzmann}).
We may now re-express this equation as
\be
\dot T\frac{dn_{\tilde{G}}}{dT}+3Hn_{\tilde{G}}=\langle \Sigmatot|v|\rangle
n^2\,,
\label{boltzmann1}
\ee
(keeping in mind that $\dot T$ passes through zero at $\Tmax$).

The cross section $\langle\Sigmatot|v|\rangle$
is given by~\cite{pradlersteffen1}
\bea
\langle\Sigmatot|v|\rangle &\equiv&
\frac{\alpha}{M^2}\nonumber\\
& =& \frac{1}{M^2} 
\frac{3\pi}{16\zeta (3)}\,
\sum_{i=1}^3
\left[ 1+
\frac{M_i^2}{3\mG^2}
\right]
\nonumber\\
&&\hspace{2.5cm}\times\, 
c_i \,g_i^2\, \ln\left(\frac{k_i}{g_i}\right)
\label{sigma-total}
\eea
where $i=1,2,3$ refers to the three gauge groups $U(1)_Y, SU(2)_L$ and 
$SU(3)_c$ respectively.  
$M=\mpl/
\sqrt{8\pi}\simeq 2.4\times 10^{18}$ GeV is the reduced Planck mass.
$M_i$ are the gaugino masses.  $g_i(T)$ are the gauge coupling constants,
while $c_i$ and $k_i$ are constants associated with the gauge groups.
$c_{1,2,3}$ are 11, 27 and 72 and $k_{1,2,3}$ are 1.266, 1.312, 1.271
respectively (Table 1 of Ref.~\cite{pradlersteffen1}).  The above
expression includes corrections to earlier expressions for the cross section
for gravitino production
in 
Refs.~\cite{BBB_npb.01} and \cite{kkm}.
Using the one loop $\beta$-function of MSSM, the
solution of the renormalization group equation for the gauge coupling
constants is given by
\be
g_i(T)\simeq \left[ g_i^{-2}(M_Z)-\frac{b_i}{8\pi^2}\ln(T/M_Z)
\right]^{-1/2}\,,
\label{coupling-const}
\ee
with $b_1=11$, $b_2=1$, $b_3=-3$.
To obtain a conservative estimate of the gravitino abundance
we take $M_i\rightarrow0$ 
as in Ref. \cite{kkm}.

\subsection{Gravitino production during reheating}
Since $\dot T$ is zero at $\Tmax$ we solve Eq. (\ref{boltzmann1}) from
$t_{\rm osc}$ to $t_{\rm max}$ and
from $t_{\rm max}$ to $t_{\rm reh}$ separately. In order to solve Eq. (\ref{boltzmann1})
we need $\Tdot$ and $H$ as functions of $T$. Eq. (\ref{rho_r}) implies that
\be
R^4 T^4 - A \Rosc^2 R^2 + A\Rosc^4=0
\label{RTeqn}\ee
where with some algebra one can show that $A=4\Tmax^4$. Then in the epoch 
$t_{\rm osc}\le t \le t_{\rm max}$ we have
\be
R^2/\Rosc^2=\frac{1-(1-4T^4/A)^\half}{2T^4/A}
\label{RT}
\ee
while for $t_{\rm max} \le t \le t_{\rm reh}$
\be
R^2/\Rosc^2=\frac{1+(1-4T^4/A)^\half}{2T^4/A}
\, .\label{RT1}
\ee
Eq. (\ref{RT1}) implies that for $t\gg\tmax$, $T\sim R^{-\half}$
(not $\sim R^{-1}$ as inflaton decay is a source of radiation).
We now define a dimensionless
variable
\be
x^2=1-\frac{4T^4}{A}=1-\frac{T^4}{T_{\rm max}^4}\,,
\label{x-def}
\ee
Then the Hubble expansion parameter $H=H_I (\Rosc/R)^2$ can be rewritten as a function of
$T$ or $x$ as
\be
H = \frac{1}{2}H_I(1+x)
\;\;\; {\rm for} \;\;\;t_{\rm osc} \le t \le t_{\rm max} \label{Hubble_tosc}
\ee
and
\be
H = \frac{1}{2}H_I(1-x)
\;\;\; {\rm for} \;\;\;t_{\rm max} \le t \le t_{\rm reh}\,.
\label{Hubble_treh}
\ee

For $\Tdot$ we differentiate $\rho_r(T) =(\pi^2/30) g_* T^4$ and $\rho_r(R)$
in Eq. (\ref{rho_r}) with respect to time, equate the results and get
\be
\dot{T}=
\frac{30}{\pi^2 g_*}\frac{1}{4T^3} \frac{d\rho_r}{dR} \dot R
=
\frac{T}{4\rho_r} \frac{d\rho_r}{dR} R H \,.
\label{Tdot}
\ee
Now using Eqs. (\ref{rho_r}), (\ref{RT},\ref{RT1}), (\ref{x-def}) and
(\ref{Hubble_tosc},\ref{Hubble_treh}) the above
equation can be recast as
\be
\dot{T}=\frac{T_{\rm max}^4 H_I}{2 T^3}x (1+x)^2
\;\;\; {\rm for} \;\;\;t_{\rm osc} \le t \le t_{\rm max}
\label{Tdot-1}
\ee
and
\be
\dot{T}=\frac{T_{\rm max}^4 H_I}{2 T^3} (-x) (1-x)^2
\;\;\; {\rm for} \;\;\;t_{\rm max} \le t \le t_{\rm reh}\,.
\label{Tdot-2}
\ee
We now solve the Boltzmann equation Eq. (\ref{boltzmann1}) in the two regimes
$t_{\rm osc} \le t \le t_{\rm max}$ and $t_{\rm max} \le t \le t_{\rm reh}$.

\subsection{Epoch: $t_{\rm osc} \le t \le t_{\rm max}$}

Using Eqs. (\ref{Hubble_tosc}) and (\ref{Tdot-1}) Eq. (\ref{boltzmann1}) can be
written as
\be
\frac{dn_{\tilde{G}}}{dx} - \frac{d_1}{(1+x)}n_{\tilde{G}} = d_2 \frac{(1-x^2)^{3/2}}{(1+x)^2}
\label{boltzman_x}
\ee
where
\be
d_1=\frac{3}{2}\;\; {\rm and} \;\;  d_2=-\frac{\alpha}{M^2} \left( \frac{\zeta(3)}{\pi^2} \right)^2
\left(\frac{ T_{\rm max}^6}{H_I} \right) \,.
\label{d1&d2}
\ee
Now we define $y=1+x$, so that Eq. (\ref{boltzman_x}) can be rewritten as
\be
\frac{dn_{\tilde{G}}}{dy}-\frac{d_1}{y}n_{\tilde{G}} =d_2 \frac{(2-y)^{3/2}}{y^{1/2}}
\label{boltzman_y}
\ee
Solving Eq. (\ref{boltzman_y}) from $y_{\rm osc}=2$ to $y$ and assuming that
$n_{\tilde{G}}(y_{\rm osc})=0$ we get the gravitino abundance as \cite{integral}
\bea
n_{\tilde{G}} (y) &=& d_2 y^{3/2}\left[\frac{2(2-y)-6}{y}\sqrt{2-y}\right. 
\nonumber\\
&& \left. - \frac{3}{\sqrt{2}}\ln \left|\frac{\sqrt{2-y}-\sqrt{2}}{\sqrt{2-y}+\sqrt{2}}
\right|\,\right]\,.
\label{abun_tosc}
\eea
Thus at $y=y_{\rm max}=1$, which corresponds to $T=T_{\rm max}$, we get the 
gravitino
abundance to be
\bea
n_{\tilde{G}}(y_{\rm max}) &=& d_2\left[-4-\frac{3}{\sqrt{2}}\ln \left|
\frac{1-\sqrt{2}}{1+\sqrt{2}} \right|\,\right]\nonumber\\
&= & -0.26 \,d_2
\label{nG_max}
\eea

\subsection{Epoch: $t_{\rm max} \le t \le t_{\rm reh} $}
Using Eqs. (\ref{Hubble_treh}) and (\ref{Tdot-2}) Eq. (\ref{boltzmann1})
can be written as 
\be
\frac{dn_{\tilde{G}}}{dx} + \frac{d_1}{(1-x)}n_{\tilde{G}} = d_2 \frac{(1-x^2)^{3/2}}{(1-x)^2}
\label{boltzman1_x}
\ee
where $d_1$ and $d_2$ are given by Eq. (\ref{d1&d2}).  Defining $y =1-x$ Eq.
(\ref{boltzman1_x}) can be rewritten as
\be
\frac{dn_{\tilde{G}}}{dy}-\frac{d_1}{y}n_{\tilde{G}} =d_2 \frac{(2-y)^{3/2}}{y^{1/2}}
\label{boltzman1_y}
\ee
Solving Eq. (\ref{boltzman1_y}) from $y_{\rm max}$ to $y$ we get
the gravitino abundance as \cite{integral}
\bea
n_{\tilde{G}} (y) &=& n_{\tilde{G}}(y_{\rm max}) \left(\frac{y}{y_{\rm max}}\right)^{3/2}
\nonumber\\
&+&d_2 y^{3/2} \left[4 + \frac{3}{\sqrt{2}}\ln \left| \frac{1-\sqrt{2}}
{1+\sqrt{2}} \right| \,\right]\nonumber\\
&+& d_2 y^{3/2} \left[ -2 \left( 1 +\frac{1}{y} \right) \sqrt{2-y}
\right. \nonumber\\
&& \left. -\frac{3}{\sqrt{2}}\ln \left|
\frac { \sqrt{2-y}-\sqrt{2}}{ \sqrt{2-y}+\sqrt{2} } \right| \,\right]
\label{gabundance_t}
\eea
Now using Eq. (\ref{nG_max}) and letting $y_{\rm max}=1$ and
$y=y_{\rm reh}$ the above equation can be written as
\bea
n_{\tilde{G}} (y_{\rm reh}) &=& d_2 \,y_{\rm reh}^{3/2} \left[ -2 \left( 1 +\frac{1}{y_{\rm reh}} \right)
\sqrt{2-y_{\rm reh}} \right. \nonumber\\
&& \left. -\frac{3}{\sqrt{2}}\ln \left|
\frac { \sqrt{2-y_{\rm reh}}-\sqrt{2}}{  \sqrt{2-y_{\rm reh}}+\sqrt{2} } 
\right| \,\right]
\label{gabundance_treh}
\eea
Using $y_{\rm reh}\approx(1/2)(T_{\rm reh}/T_{\rm max})^4$  for
$\Treh\ll\Tmax$
we get the gravitino abundance at
$T_{\rm reh}$ as
\bea
n_{\tilde{G}}(T_{\rm reh}) &=&  \frac{d_2}{2^{3/2}}(T_{\rm reh}/T_{\rm max})^6\nonumber\\
&&\left[-4\sqrt{2} \left( \frac{T_{\rm max}} {T_{\rm reh}} \right)^4 \left(1 + \frac{3}{8}
\frac{T_{\rm reh}^4}{T_{\rm max}^4}-\frac{1}{16} \frac{T_{\rm reh}^8}{T_{\rm max}^8} \right)
\right. \nonumber\\
&-& \left. \frac{3}{\sqrt{2}} \ln \left(   \frac{8 T_{\rm reh}^4}{16 T_{\rm max}^4-T_{\rm reh}^4}
\right) \right]
\eea
For $\Treh\ll\Tmax$
the dominant contribution comes from the $(T_{\rm max}/T_{\rm reh})^4$ term
in the square brackets. Then we can approximate
the gravitational abundance at $\Treh$ as
\bea
n_{\tilde{G}}(T_{\rm reh}) &\simeq & \frac{d_2}{2^{3/2}}(T_{\rm reh}/T_{\rm max})^6
\left[-4\sqrt{2} \left( \frac{T_{\rm max}}{T_{\rm reh}} \right)^4\right] 
\nonumber\\
&=& 2 \left( \frac{\alpha}{M^2} \right) \left( \frac{\zeta(3) }{\pi^2}\right)^2 \frac{T_{\rm max}^4}{H_I}
T_{\rm reh}^2\,.
\label{gabundance_final}
\eea

\begin{figure}
\begin{center}
\epsfig{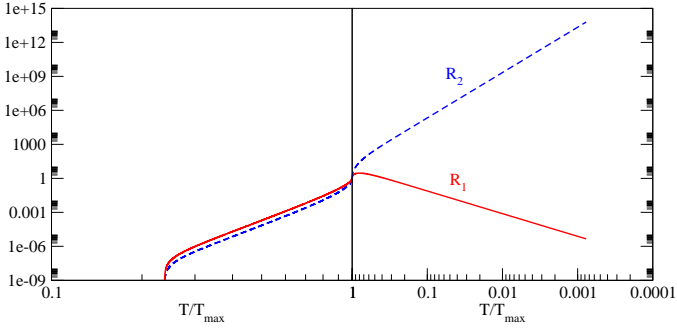}
\caption{
The production of gravitinos is shown from $t_{\rm osc}$ to 
$t_{\rm reh}$ through
$t_{\rm max}$ as a function of temperature for values mentioned
in the Discussion.  
The temperature rises from 0 at $\tosc$ to $\Tmax$ at $\tmax$ and then
falls to $\Treh$ at $\treh$.  
The (normalised) gravitino number density ${\rm R_1}=\nG(T)/\nG(\Tmax)$ 
rises from 0 to its maximum value at $\tMAX>\tmax$, when $\TMAX\simeq0.8\Tmax$, 
and then decreases till 
$\treh$ [red solid line]. The (normalised) gravitino number density per comoving volume 
${\rm R_2}=\bar n_G(T)/\bar n_G(\Tmax)$ is also plotted [blue dashed line].
\label{fig-1}}
\end{center}
\end{figure}

\subsection{Gravitino production in the radiation dominated era}
After the inflaton field decays at $\treh$ the universe
enters the radiation dominated era. Unlike the reheating era during
which the entropy continuously increases, in the radiation dominated
era the total entropy remain constant (except for epochs of
out-of-equilibrium decays). Therefore it is useful to express the
abundance of any species $i$ as $Y_i=n_i/s$, where $n_i$ is
the number density of the species $i$ in a physical volume
and $s$ is the entropy density given by
\be
s=\frac{2\pi^2}{45}g_*T^3\,.
\label{entropy-density}
\ee
We take $g_*=228.75$ in the MSSM for the temperature range of interest.
One can now re-express Eq. (\ref{boltzmann1})
as
\be
\dot T\frac{d\YG}{dT}=\langle \Sigmatot|v|\rangle
 Y n \,.
\label{boltzmann2}
\ee
To obtain $\Tdot$ we use the temperature-time relation for
the radiation dominated era, namely,
\be
T={\Treh}\frac{1}{\left[2\Hreh(t-\treh)+1\right]^\frac{1}{2}}\,,
\label{T-t-reh}\ee
where
\be
\Hreh= \sqrt{ \frac {8\pi^3 g_{*\rm{reh}}} {90} }\frac{\Treh^2}{\mpl}\,.
\label{H-reh-value}
\ee
(For $t\gg\treh$, $T\sim t^{-\frac 1 2}$.)
Therefore $\dot T$ is given by
\begin{equation}
\dot T =-\frac{\Hreh}{\Treh^2}T^3 =
-\left( \frac{g_{*\rm{reh}}\pi^2}{90}\right)^{\frac{1}{2}}\frac{T^3}{M}\,.
\label{Tdot-3}\end{equation}
Then
\be
\frac{d\YG}{dT}=-\left( \frac{90}{\greh\pi^2} \right)^{1/2}
\left(\frac{45}{2\pi^2 g_*}\right) \left(\frac{\alpha}{M}\right)
\left( \frac{\zeta(3)}{\pi^2} \right)^2\,.
\label{YT-boltzmann}
\ee
Assuming $\alpha$ to be independent of temperature and integrating
the above equation from $\Treh$ to $\Tf$, the final temperature,
we get the total gravitino abundance at $\Tf$ to be
\be
Y_{\tilde{G}}(\Tf) = Y_{\tilde{G}}(\Treh)+Y^{\rm rad}_{\tilde{G}}(T_{f})
\label{total_abundance}
\ee
where
\bea
Y_{\tilde{G}}^{\rm rad}(T_f) &=& \left( \frac{90}{\greh\pi^2}
\right)^{1/2}\left(\frac{45}{2\pi^2 \greh}\right)\nonumber\\
&&\times  \left(
\frac{\alpha}{M}\right)\left( \frac{\zeta(3)}{\pi^2} \right)^2 (\Treh-\Tf)
\label{gravitino-density}
\eea
is the gravitino abundance produced in the radiation dominated era.
We have used $\greh$ in
the expression for $Y_{\tilde{G}}^{\rm rad}$ and
ignored the variation of $g_*$ with temperature.  This is justified
since most of the gravitinos are generated close to $\Treh$.
Using Eqs. (\ref{gabundance_final}) and (\ref{entropy-density})
\be
Y_{\tilde{G}}(\Treh)=\frac{2\alpha}{M^2}\left( \frac{\zeta(3)}
{\pi^2} \right)^2\left(\frac{45}{2\pi^2 \greh}\right)
\frac{\Tmax^4}{H_I \Treh}\,.
\label{density-treh}
\ee
This contribution is usually neglected while obtaining the gravitino bound.
As we see below this is comparable with the second term in Eq.
(\ref{total_abundance}).  Using Eqs. (\ref{gravitino-density}) and
(\ref{density-treh}) in Eq. (\ref{total_abundance}) we get
\bea
Y_{\tilde{G}}(\Tf) &=& \frac{\alpha}{M^2}
\left( \frac{\zeta(3)}{\pi^2} \right)^2
\left(\frac{45}{2\pi^2 \greh}\right) \left[2\frac{\Tmax^4}
{H_I \Treh}\right.\nonumber\\
&&+ \left. M\left( \frac{90}{\greh\pi^2} \right)^{1/2} \Treh\right]\,,
\label{comparison}
\eea
where we have used $\Tf\ll\Treh$. Relating $\Tmax$ to $\Treh$ from Eqs.
(\ref{HI-value}), (\ref{T-max}) and (\ref{reheat-temp}) 
the total gravitino abundance
is then given by
\bea
Y_{\tilde{G}}(T_f) &=& \frac{3 \alpha \Treh}{M}
\left( \frac{\zeta(3)}{\pi^2} \right)^2\left( \frac{45}
{2\pi^2 \greh^{3/2}}\right)\nonumber\\
&& \left[ 0.49 + 1.0 \right]\,,
\label{comparison-value}
\eea
where we have used $\greh$ in the expressions for $\Tmax$.
\begin{figure}
\begin{center}
\epsfig{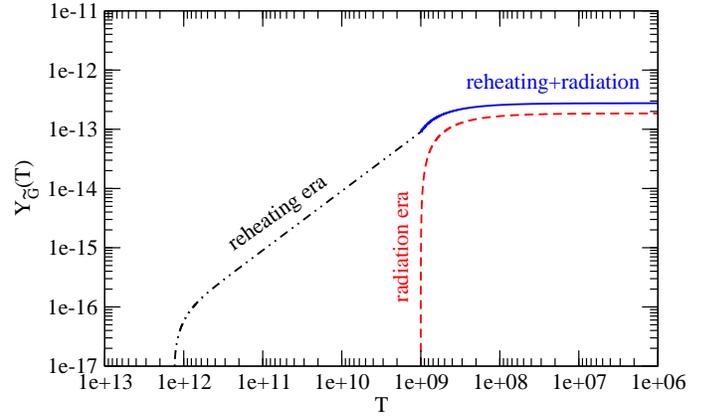}
\caption{
$\YG=\nG/s$ generated during the reheating era,
the radiation dominated
era, and the sum of contributions from both eras
are shown as a function of the temperature $T$ for
$t>\tmax$.
$\Treh$ and $M_I$
are chosen
to be $10^9\gev$ and $10^{16}\gev$ respectively,
and so $\Tmax\approx1.3\times10^{12}\gev$.  
Since $\YG$ in both eras is largely
generated close to $\Treh$,
$\alpha$ is evaluated at $\Treh$.
The final value of $\YG$ is $\approx 3\times 10^{-13}$.
\label{fig-2}}
\end{center}
\end{figure}

\section{Discussion}

The detailed dynamics of
gravitino
production 
from $t_{\rm osc}$ to $t_{\rm reh}$ is shown in Fig. (\ref{fig-1}) 
as a function of
the temperature  
using Eqs. (\ref{abun_tosc}) and (\ref{gabundance_t}).
We normalise the gravitino number density with respect
to the value at $\Tmax$.
$\Treh$ and $M_I$
are chosen
to be $10^9\gev$ and $10^{16}\gev$ respectively,
and so $\Tmax\approx1.3\times10^{12}\gev$.
$\alpha$ is treated as constant and evaluated at $\Treh$ since most
gravitinos are produced near $\Treh$.  $\alpha$ is 15.1,
using $g_i(M_z)$ obtained from
$\alpha_{EM}(M_Z)=1/128$, $\sin^2\theta_W(M_Z)=0.231$, $\alpha_s(M_Z)=0.119$,
and $M_Z=91.2\gev$ \cite{pdg}.

It can be seen in Fig. (\ref{fig-1}) that during reheating
the number density of gravitinos monotonically
increases from
$t_{\rm osc}$ to a time $t_{\rm MAX} > \tmax $, where $t_{\rm MAX}$
corresponds to
a temperature $\TMAX \simeq 0.8\,\Tmax$ and $\nG(\tMAX)\approx3\,
\nG(\tmax)$.  Subsequently it decreases till $\treh$ to a value
$7.7 \,(\Treh/\Tmax)^2 \,\nG(\tmax)$.
Taking the number density per comoving volume as
$\bar n_G (T) = (\nG R^3)/\Rosc^3$
we also plot $\bar n_G$
(normalised to the value at $\Tmax$) by
using Eqs. (\ref{RT}), (\ref{RT1}), (\ref{abun_tosc}) and (\ref{gabundance_t}).
From the plot of $\bar n_G (T)$ in Fig. (\ref{fig-1})
it can be seen that
most gravitinos during reheating are produced close to $T_{\rm reh}$.

In Fig. (\ref{fig-2}) we show the contribution to the gravitino abundance
from the reheating era and from the subsequent radiation dominated era,
and the sum of these contributions.
From Eq. (\ref{comparison-value}) it is clear that the gravitino production 
during the
reheating era is almost half of that during the radiation dominated era
even though a priori one would
not have expected the gravitino production in both these eras to be similar.  
While the gravitino
abundance generated during reheating is a function of $\Tmax$ it is 
interesting that
it can be re-expressed as independent of
$\Tmax$, and as a function of only $\Treh$. 
Moreover, the contribution to $\YG$ from the
reheating era is linearly proportional to $\Treh$, as it is for the radiation 
dominated
era.  These results are similar to those obtained in Ref.~\cite{rangarajansahu}.
The linear dependence on $\Treh$ makes it simple to revise the constraints on
$\Treh$ based on the upper limit on the gravitino abundance - the upper bound on 
$\Treh$ is lowered by
a factor of 3/2.  Since $\Tmax\propto\sqrt\Treh$, $\Tmax$ is not 
affected much.  Therefore
models of leptogenesis that invoke a large $\Tmax$ to create heavy Majorana 
neutrinos are
not significantly impacted.

Above we partly ignored inflaton decay in our analysis, i.e., we did not 
include the effect of
$\exp[-\Gamma_\phi (t-\tosc)]$ 
in $\rho_\phi$ in Eq. (\ref{rhoreqn}).  One might be
concerned that this will lead to inaccuracies close to $\treh$ when most
of the gravitinos are produced.  However if one writes $\rho_\phi \sim R^{-4}
\exp(-\Gamma_\phi t)\sim t^{-2} \exp(-\Gamma_\phi t)$ for $t\gg\tmax$ then 
$\dot\rho_\phi/\rho_\phi= -2/t -\Gamma_\phi$.  Therefore even till close to 
$\treh=\Gamma_\phi^{-1}$ $\rho_\phi$ decreases primarily due to the expansion
of the universe.  Furthermore, 
if we
follow the value of $\YG=\nG/s$ we find (from Eq. (\ref{density-treh}))
that near $\treh$ it increases as
$T^{-1}\sim R^\half\sim t^{\frac 14}$.
At $t \approx 0.1 \Gamma_\phi^{-1}$  56\% of our estimate of
$\YG$ is already generated while
decay has led to a reduction in $\rho_\phi$ of only 9\%.  
Keeping in mind the above arguments, 
we expect that
the error in our estimate of $\YG$ will 
not be large.
A more accurate estimate will require a numerical analysis.
(For the quadratic potential, Ref.~\cite{rangarajansahu} (analytic)
and Ref.~\cite{kkm} (numerical)
obtained a gravitino abundance of $1.9\times10^{-13}$ and $1.5\times10^{-13}$
respectively 
for $\Treh$, as defined in Ref.~\cite{rangarajansahu}, set to $10^9 \gev$.
\footnote{
Refs.~\cite{rangarajansahu}
and~\cite{kkm} define the reheat temperature differently.
Note that both their results are enhanced by 27\% if one uses the gravitino 
production rate from Ref.~\cite{pradlersteffen1}.}) 

\section{Conclusion}

In conclusion, in this article we have calculated the
gravitino abundance generated during reheating for an inflationary model with a quartic
potential
during reheating.
We find that the gravitino abundance generated
during reheating
is a function of the largest temperature during reheating.  However it can be re-expressed
in terms of
the reheat temperature only and we
find that it is linearly proportional to the the reheat temperature,
as in the standard calculation of gravitinos produced in the radiation dominated
era after reheating. Furthermore, we find that this abundance is 49\% of
the abundance of gravitinos
generated in the radiation dominated era.  This lowers the upper bound on the
reheat temperature
by a factor of 3/2.  However this does not significant change the
viability of leptogenesis scenarios.

{\bf Acknowledgement:}
NS would like to thank Anupam Mazumdar and Kazunori Kohri for useful 
discussions.
NS was supported by the European Union through the Marie Curie Research and
Training Network ``UniverseNet" (MRTN-CT-2006-035863).

%

\end{document}